\documentclass[twocolumn,prl,showpacs,superscriptaddress]{revtex4}
\usepackage{graphicx}
\usepackage{amsmath}
\usepackage{amssymb}

\begin{document}
\title{Continuity of the Explosive Percolation Transition}

\author{Hyun Keun Lee}
\affiliation{Department of Physics, University of Seoul, Seoul 130-743, Korea}
\affiliation{BK21 Physics Research Division, Sungkyunkwan University, Suwon 440-746, Korea}
\author{Beom Jun Kim}
\affiliation{BK21 Physics Research Division, Sungkyunkwan University, Suwon 440-746, Korea}
\author{Hyunggyu Park}
\affiliation{School of Physics, Korea Institute for Advanced Study,
Seoul 130-722, Korea}

\date{\today}

\begin{abstract}
The explosive percolation problem on the complete graph is investigated via extensive
numerical simulations. We obtain the cluster-size distribution at the moment
when the cluster size heterogeneity becomes maximum.
The distribution is found to be well described by the power-law form with
the decay exponent $\tau = 2.06(2)$, followed by a hump.
We then use the finite-size scaling method to make all the distributions
at various system sizes up to $N=2^{37}$ collapse perfectly onto a scaling curve
characterized solely by the single exponent $\tau$.
We also observe that the instant of that collapse converges to a
well-defined percolation threshold from below as $N\rightarrow\infty$.
Based on these observations, we show that the explosive percolation transition
in the model should be continuous, contrary to the widely-spread belief of its discontinuity.

\end{abstract}


\maketitle

The term {\it explosive percolation} was proposed in
Ref.~\cite{AchlioptasSCIENCE2009} to describe
a sudden appearance of a  macroscopic cluster in a
network growth model with the so-called product rule considered on the complete graph.
This growth rule, named as the Achlioptas process (AP),
is then studied on
the two-dimensional lattice~\cite{ZiffPRL2009,ZiffPRE2010} and on the
scale-free networks~\cite{RadicchiPRL2009,RadicchiPRE2010,KimPRL2009} as well,
yielding similar results.
That suddenness has been widely believed to indicate a
discontinuity at the percolation transition in the thermodynamic
limit~\cite{MoreiraPRE2010,KimPRE2010},
and the similar explosiveness has been
observed with the other growth rules proposed
later~\cite{Manna,FriedmanPRL2009,KimPREr2010,DSouzaPRL2010,HerrmannPRL2010}.
These observations of the explosiveness
have drawn much interest due to the striking difference
from the well-known continuous transition
in the standard percolation models~\cite{Christensen}.
However, in our point of view, the explosiveness has not been carefully investigated as yet
enough to draw a decisive conclusion on the discontinuity, and possibly just represents
an extremely steep but still continuous transition.

Friedman and Landsberg~\cite{FriedmanPRL2009} have suggested
the argument of the {\it powder keg} as a circumstantial description
to explain the apparent discontinuity of the explosive percolation transition.
Meanwhile, da Costa {\it et al.}~\cite{Costa}
have reported that the explosive percolation is actually continuous for a modified
version of the AP by analytically deriving the critical scaling relations based on
numerical observations of power-law critical distribution of cluster size~\cite{RW}.
In this Letter, we try to unmask the (dis)continuity in a systematic and direct way
by performing a careful finite-size-scaling analysis
at newly introduced {\em pseudo}-transition points for finite systems
and show that
the explosive percolation transition on the complete graph is indeed continuous
in the thermodynamic limit.

The model we study is the AP with the product rule on the complete graph~\cite{AchlioptasSCIENCE2009}. Start with $N$ nodes with all links unoccupied. At each step, choose two possible unoccupied links randomly between nodes. Then, select the link merging two clusters with a smaller product of the two cluster sizes.
Here, a cluster is defined as a set of nodes connected each other via occupied links.
This procedure is repeated until all nodes are connected as a whole.
The number of occupied links $L$ increases one by one at each step and
the occupied link density (or {\em time}), $t=L/N$, serves as the control parameter for the model.
The interested observable is the largest cluster size $G(t)$ which becomes macroscopic (linear in $N$) at sufficiently large $t$. The order parameter is defined as the relative size of the largest cluster, $g(t)=G(t)/N$, which remains at zero below the threshold $t_c$ and becomes finite
for $t>t_c$ in the $N=\infty$ limit.

The main question is whether the gap, $g(t)|_{t\rightarrow t_c^+}$, vanishes (continuous
transition) or approaches a non-zero constant (finite jump).
It may be natural to use the information above the transition point ($t>t_c)$ in order to
prove the (non-)existence of the gap or estimate the gap size. Thus, most of previous studies
have focussed on this information~\cite{AchlioptasSCIENCE2009,ZiffPRL2009,ZiffPRE2010,RadicchiPRL2009,RadicchiPRE2010,KimPRL2009,MoreiraPRE2010,KimPRE2010,FriedmanPRL2009,Manna,KimPREr2010,DSouzaPRL2010,HerrmannPRL2010,Costa},
but could not provide a definitive answer due to
the extremely slow convergence of the order parameter in system size.
In this work, we took the opposite approach. Using the accurate information below $t_c$,
it is still possible to derive the upper bound for the gap, which turns out to vanish
as $N\rightarrow \infty$. This guarantees the vanishing gap at the transition.

Our strategy is as follows: (i) Set up lower and upper pseudo-transition points, $t_l(N)$ and $t_u(N)$,  for finite size $N$ below and above the true asymptotic percolation transition point $t_c$, respectively. We expect that both pseudo-transition points converge to $t_c$ as $N\rightarrow\infty$.
(ii) Find the upper bound for the size increase of a largest cluster
$\Delta G$ between $t_l(N)$ and $t_u(N)$.  (iii) Show that this upper bound is sublinear in $N$, which implies no macroscopic
jump of the largest cluster size through the percolation transition.
This completes the proof of
the continuity at the explosive percolation transition. All procedures are done via extensive numerical simulations, typically up to $N=2^{37}\approx 1.37\times 10^{11}$ and the
average is done over $100\sim 5000$ runs.

\begin{figure}
\includegraphics*[width=\columnwidth]{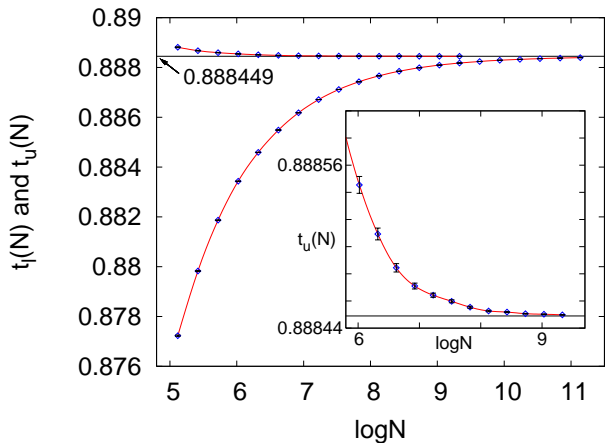}
\caption{(Color online) Convergence of $t_{l}$ (the lower branch) and
$t_{u}$ (the upper one) averaged for $100\sim 5000$ different realizations for each
system size ($N=2^{17},2^{18},..,2^{37}$). Errors are smaller than symbol sizes,
if not shown explicitly.
Lines  are just guides to the eyes.
As $N$ increases, $t_u$ ($t_l$) decreases
(increases), approaching to the well-defined value in the
thermodynamic limit displayed as the horizontal line.
For clarity, the upper branch is vertically enlarged in the inset.}
\label{fig1}
\end{figure}

The most crucial step is to define the two pseudo-transition points at the microscopic step level.
First, we introduce the lower pseudo-transition point $t_l(N)$ as the instant when the
{\it cluster size heterogeneity} (the number of distinct cluster sizes)
becomes maximum. For small $t$, the cluster size heterogeneity increases with $t$ due to the randomness of clustering processes. However, the emergence of a macroscopic percolating cluster which continuously absorbs small clusters causes the heterogeneity to decrease and eventually the whole system becomes one  cluster. Due to the mechanism of suppressing
the emergence of large clusters,
one may argue that the heterogeneity increases slowly but steadily up to just before
the explosion when many different size clusters merge into one big macroscopic cluster.
Thus it is reasonable to consider the {\em maximum heterogeneity} as a preceding symptom of the percolating onset for finite systems. In Fig.~\ref{fig1}, the average values of $t_l(N)$ are plotted against $N$ (lower branch), which
converge to the asymptotic value of $t_c=0.8884490(5)$ from below, as expected.

\begin{figure}
\includegraphics*[width=\columnwidth]{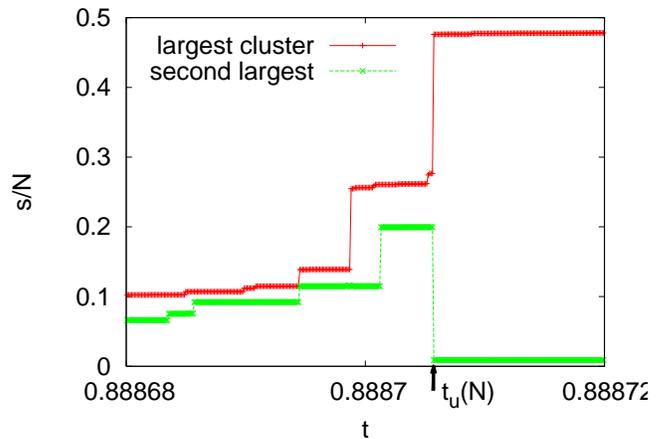}
\caption{(Color online) Evolutions of the largest cluster size (the upper
curve) and the size of the second largest cluster (the lower curve)
versus the growth step $t$. As $t$ crosses $t_u$ from below,
the size of the largest  cluster exhibits a sudden biggest increase
since the maximum second-largest cluster merges into it at that moment.
The system size is $N=2^{23}$.
}
\label{fig2}
\end{figure}

Second, we expect that the growth rate of the largest cluster also becomes maximum at the percolation transition. Microscopically, the upper pseudo-transition point $t_u(N)$ is defined as
one step after the moment when the second-largest cluster size becomes maximum.
Thus $G(t)$ can experience
a largest increase exactly when $t$ exceeds $t_u$, since the second-largest cluster
merges into the largest cluster. A typical growing process is displayed in Fig.~\ref{fig2}.
Note that the second-largest cluster never recovers its size after merging into the largest
cluster. So there will be no explosive increase of the largest cluster size for $t>t_u(N)$.
Dominance of one percolating cluster is the characteristic of the percolating phase.
So it is reasonable to expect that $t_u(N)$ is just above $t_c$, which is consistent with  numerical results
(see the inset in Fig.~\ref{fig1}).
The average values of $t_u(N)$  converge to the same
asymptotic value of $t_c$ from above as $N\rightarrow\infty$.

The sample-to-sample fluctuations decrease with $N^{-0.5}$ (not shown here),
which implies that both $t_u$ and $t_l$ are self-averaging~\cite{AH},
so not only the critical point but also any sample-averaged quantity
are well defined in the asymptotic limit~\cite{RW}.
We also find numerically
\begin{equation}
t_{u}(N)-t_{l}(N) \sim N^{-\delta},
\label{th-t}
\end{equation}
with $\delta=0.39(3)$~\cite{explan0}.

\begin{figure}
\includegraphics*[width=\columnwidth]{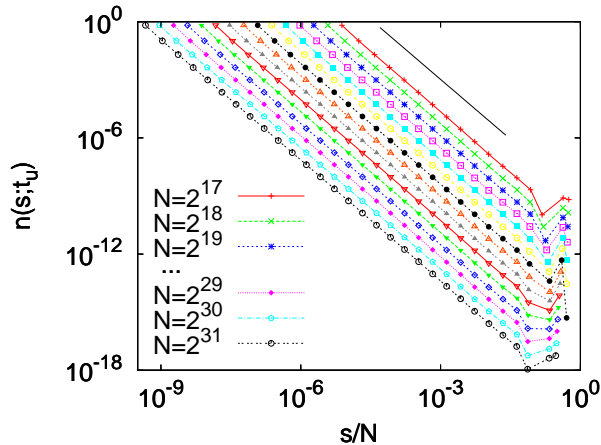}
\caption{(Color online)
Log-binned cluster-size distributions at $t_u$ for each $N$,
where the horizontal axis is the cluster size $s$ divided by $N$ for convenience.
The solid line above is a guiding line of which the decay exponent is 2.06.}
\label{fig3}
\end{figure}

Now we consider $\Delta g=g(t_u)-g(t_l)=\Delta G/N$,
which is the growth of the largest cluster density between two pseudo-transition points
through the asymptotic transition point $t_c$. In the thermodynamic limit,
it will be the jump size (if any) of the order parameter at the percolation transition.
As $g(t_l)$ must vanish as $N\rightarrow\infty$,
we only need the information of $g(t_u)$ in principle to calculate
$ \Delta g|_{N\rightarrow\infty} $.
Figure \ref{fig3} shows the cluster-size distribution $n(s;t_u)$ of cluster size $s$
(normalized by the total
number of clusters $C$) at $t_u(N)$ for various different sizes $N$.
The distribution  fits extremely well with a power-law form,
 $n(s;t_u)\sim s^{-\tau}$ with the decay exponent $\tau=2.06(2)$ in a huge range, which is then accompanied by a little dip in the end.
In Fig.~\ref{fig3}, the largest cluster size $G(t_u)$ depending on
the dip structure near the upper cutoff shows a slight trend of the sublinearity in $N$ (moving left in the axis
of $s/N$ as $N$ increases), which may be one symptom for the continuous transition. However,
as discussed before, it can not be conclusive even with huge system sizes studied here.

If one assumes a conventional {\em natural} cutoff of the power-law type distribution function,
the upper cutoff which
should be proportional to the largest cluster size $G(t_u)$ will scale as
$N^{1/(\tau-1)}\simeq N^{0.94}$ with $\tau\simeq 2.06$.
Sublinearity is estimated only by $6\%$, which may call for
a huge system size like $N\sim 10^{17}$ (beyond the present computing capability)
to reach a reasonable scaling regime
($g(t_u)\lesssim 0.1$) and get any sensible extrapolation to the thermodynamic limit. Most of previous
studies~\cite{AchlioptasSCIENCE2009,ZiffPRL2009,ZiffPRE2010,RadicchiPRL2009,RadicchiPRE2010,KimPRL2009,MoreiraPRE2010,KimPRE2010,FriedmanPRL2009,Manna,KimPREr2010,DSouzaPRL2010,HerrmannPRL2010,Costa} basically depend on the data in this
supercritical regime ($t>t_c$).
Nevertheless, the scaling plot with this natural
cutoff shows a {\em reasonable} collapse including the dip structure at the end, but
involving big statistical errors (not shown here).

\begin{figure}
\includegraphics*[width=\columnwidth]{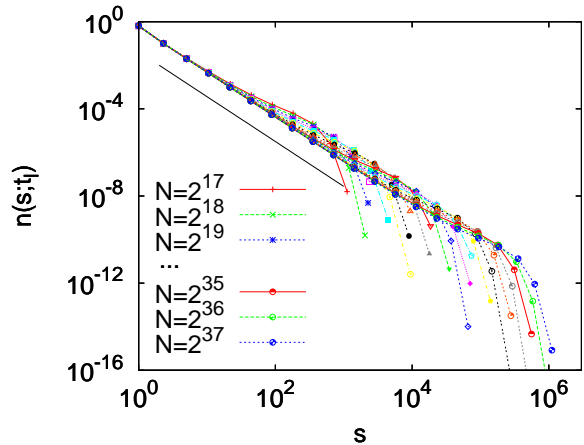}
\caption{(Color online)
Log-binned cluster-size distributions at $t_l$ for each $N$.
The decay exponent of the guiding solid line below is 2.06.}
\label{fig4}
\end{figure}

\begin{figure}
\includegraphics*[width=\columnwidth]{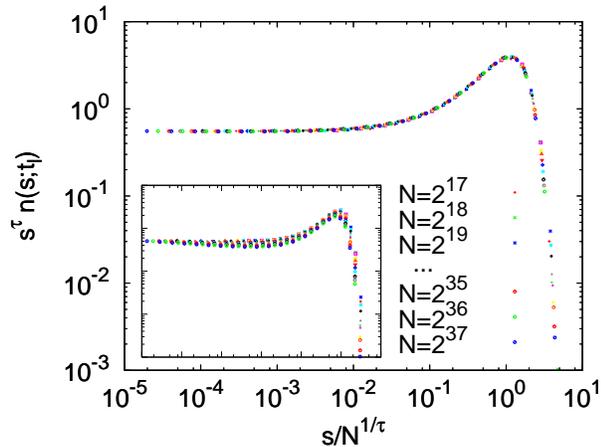}
\caption{(Color online) Finite-size scaling of the cluster
size distribution $n(s)$ at $t_{l}(N)$: All data points plotted in the
form of $s^\tau n(s)$ versus $s N^{-1/\tau}$ at $\tau=2.06$
 collapse into a single curve.
Inset: The same plot but with $\tau=2$. This clearly shows
that $\tau \approx 2.06$ ($>2$) gives us the much better quality of the FSS
collapse.}
\label{fig5}
\end{figure}

In efforts to find conclusive evidences, we scrutinize the cluster-size distribution $n(s;t_l)$
at $t_l(N)$, which shows again the power-law decay with the same
decay exponent $\tau=2.06(2)$ followed by a hump near the upper cutoff (see Fig.~\ref{fig4}).
It has a much shorter (but still quite broad) power-law regime, but exhibits much better statistics even in the hump region (see Fig~\ref{fig5}).
This power-law scaling with a hump-like structure at the end has been also reported in previous
studies~\cite{ZiffPRE2010,Manna,KimPRE2010,Costa,RadicchiPRE2010}.

In contrast to the cluster distribution at $t_u(N)$, $n(s;t_l)$ shows a fast exponential decay
near the cutoff $s_f$. This sharp cutoff originates from the nature of the growth (product) rule
which discourages the merging of bigger clusters before explosion.
One may estimate $s_f(N)$ at $t_l$ as follows. It is appropriate to estimate
the upper cutoff $s_f$ by assuming the $O(1)$ number of clusters left beyond the cutoff,
i.e.~$\sum_{s \ge s_{f}} n(s) \sim 1/C$, where $C$ is the total
number of clusters in the system. Note that $C$ scales linearly with $N$~\cite{explan2}.
Since $n(s)$ decays exponentially fast (or faster) near
$s_{f}$, $\sum_{s \ge s_{f}} n(s) \approx (\Delta s_f) n(s_f)$
with a finite characteristic scale $\Delta s_f$ of the fast decaying part.
Consequently, one can find $n({s_{f}}) \sim s_{f}^{-\tau} \sim 1/C \sim 1/N$, which
leads to
\begin{equation}
s_{f}(N) \sim N^{1/\tau},
\label{sf}
\end{equation}
at $t=t_l(N)$. Note that this cutoff scales differently from the
natural cutoff.
The huge difference in the largest cluster size $G(t)$ just below and above $t_c$
leads to its abrupt and explosive increase through the percolation transition,
which is the main difference between the ordinary and explosive percolation.
However, the magnitude of the explosion may be still sublinear in $N$ as discussed before.

The validity of Eq.~(\ref{sf}) can be checked numerically through the finite-size-scaling (FSS)
analysis for the entire distribution function $n(s;t_l)$.
Assuming a {\em single} characteristic cluster size (proportional to $s_f$),
one can write a FSS form for the distribution $n(s;t_l)$ as
\begin {equation}
n(s;t_l) =
s^{-\tau}f(s/s_f) = s^{-\tau}f(s N^{-1/\tau}),
\label{FSS}
\end{equation}
where $f(x)$ is the scaling function that becomes an $O(1)$ constant for $x \ll 1$
and decays exponentially fast (or faster) for $x\gtrsim 1$.
In Fig.~\ref{fig5}, we plot $s^{\tau}n(s;t_l)$ averaged
over 100 different runs versus the scaling variable $s N^{-1/\tau}$
for all 21 different system sizes $N=2^{17}, 2^{18}, \cdots, 2^{36},2^{37}$,
using $\tau=2.06$.
As is clearly seen, the collapse of the data points is {\em perfect} including
both the power-law decay part and also the hump structure near the end.
This remarkable collapse validates the single-variable FSS form
of Eq.~(\ref{FSS}) without any doubt. Therefore, we now have
the most precise and full information on the cluster size distribution
just below the transition for large $N$.
Comparison with the scaling collapse
plot using $\tau = 2$ (see the inset of Fig.~\ref{fig5}) leads to the
definite conclusion that $\tau$ must be larger than 2.

Now we are ready to derive the upper bound of $\Delta g=g(t_u)-g(t_l)$.
From $t=t_l$ to $t=t_u$, we need $\Delta L=N(t_u-t_l)$ steps ($\Delta L$ links added).
One may imagine the {\em ideal} process to maximize the growth of the largest cluster
$G(t)$, starting from the well-known cluster distribution  $n(s;t_l)$ at $t_l$,
by adding $\Delta L$ links one by one. This ideal process can be easily implemented
by simply linking and merging the largest cluster with the next largest cluster at each step
and repeating it till all $\Delta L$ links are exhausted. Then, all clusters of
size $s>s_\delta$  will merge into one cluster, which becomes the largest cluster
after $\Delta L$ steps. The threshold value $s_\delta$ is determined by
balancing the total number of merged clusters with the total number of links added;
$C\sum_{s>s_\delta} n(s;t_l)=\Delta L$ with $C$ the total number of clusters at $t=t_l$.

During this ideal process, the largest cluster $G(t)$ grows by
the amount of $C\sum_{s>s_{\delta}} s n(s;t_l)$.
One can easily estimate $s_\delta\sim (C/\Delta L)^{1/(1-\tau)}\sim N^{\delta/(\tau-1)}$, using the single-variable FSS form of $n(s;t_l)$ of Eq.~(\ref{FSS}) with Eq.~(\ref{th-t}).
Finally we get the strict upper bound for $\Delta g$ as
\begin{equation}
\Delta g \lesssim  s_\delta^{2-\tau}\sim N^{-\delta(\tau-2)/(\tau-1)}\approx N^{-0.022}.
\label{cont}
\end{equation}
This shows that the order parameter jump $\Delta g$ at the percolation transition
should vanish as the $N\rightarrow\infty$ limit, if  $\delta>0$ and $\tau>2$,
which are undoubtedly confirmed in our numerical simulations.
Therefore, we conclude that the explosive percolation transition is indeed {\em continuous}.

In summary, we showed that the explosive percolation transition on the complete graph is
continuous by exploiting the high-precision cluster-size information
at the moment of the maximum
cluster heterogeneity, $t_l(N)$, approaching the asymptotic transition point $t_c$
from below.
The cluster-size distribution  displays the power-law scaling
with the decay exponent $\tau=2.06(2)$, followed by
a hump with a sharp cutoff $s_f\sim N^{1/\tau}$.
It is explicitly shown that
the existence of the single-variable finite-size scaling  at $t_l(N)$ solely
guarantees the continuity of the transition if
$\tau>2$. Therefore, the scaling and the discontinuity
can not be compatible near the explosive percolation transition as in usual
critical phenomena. The explosiveness originates from the huge difference
in the largest-cluster-size scaling in $N$ below and above the transition.
However, it is not enough to invoke a discontinuity at the transition.

Our approach can be applied to many other models including
various different types of explosive percolation models to
clarify the (dis)continuity. Applications to other
explosive percolation problems and also the low-dimensional cases are
currently under investigation.

{\em Note added.} A few days before our submission, Grassberger {et al.}~posted
a preprint~\cite{Grass} drawing a similar conclusion, but following a completely
different approach.

This work was supported by Basic Science Research Program
through NRF grants (No.2010-0008758(BJK) and No.2010-0009697(HP))
funded by the MEST.


\begin{thebibliography}{99}
\bibitem{explan0} In fact, $|t_{(u,l)}(N)-t_c| \sim N^{-{\delta_{(u,l)}}}$ with $\delta_u = 0.64(2)$
and $\delta_l=0.39(3)$. So the exponent $\delta$ is determined soley by $t_l(N)$.
\end{thebibliography}
\end{document}